\documentclass{article}
\usepackage{spconf,amsmath,graphicx}
\usepackage{enumitem,kantlipsum}
\usepackage{pgfplots}
\usepackage{tikz}
\usepackage{wrapfig}

\usetikzlibrary{matrix}
\definecolor{bblue}{HTML}{4F81BD}
\definecolor{rred}{HTML}{C0504D}
\definecolor{ggreen}{HTML}{9BBB59}
\definecolor{ppurple}{HTML}{9F4C7C}
\tikzset{ 
    table/.style={
        matrix of nodes,
        row sep=-\pgflinewidth,
        column sep=-\pgflinewidth,
        nodes={
            rectangle,
            draw=black,
            align=center
        },
        minimum height=1em,
        text depth=0ex,
        text height=1ex,
        nodes in empty cells,
        row 4/.style={
            nodes={fill=purple!20}
        },
        row 8/.style={
            nodes={fill=purple!20}
        },
        row 13/.style={
            nodes={fill=blue!20}
        },
        row 14/.style={
            nodes={fill=blue!20}
        },
        row 15/.style={
            nodes={fill=gray!20}
        },
        row 16/.style={
            nodes={fill=gray!20}
        },
        row 17/.style={
            nodes={fill=yellow!}
        },
        column 1/.style={
            nodes={text width=4em,font=\bfseries}
        },
        column 3/.style={
            nodes={text width=4em}
        },
        row 1/.style={
            nodes={
                fill=black,
                text=white,
                font=\bfseries
            }
        }
    }
}

\title{A Comparison Of Deep Learning Methods for Environmental Sound Detection}
\name{Juncheng Li*, Wei Dai*, Florian Metze*, Shuhui Qu, and Samarjit Das}
\address{\{junchenl,wdai,fmetze\}@cs.cmu.edu, shuhuiq@stanford.edu, samarjit.das@us.bosch.com}
\vspace{-0.5cm}

\begin{document}
%
\maketitle

\begin{abstract}
\vspace{-0.1cm}
Environmental sound detection is a challenging application of machine learning because of the noisy nature of the signal, and the small amount of (labeled) data that is typically available. 
This work thus presents a comparison of several state-of-the-art Deep Learning models on the IEEE challenge on Detection and Classification of Acoustic Scenes and Events (DCASE) 2016 challenge task and data, classifying sounds into one of fifteen common indoor and outdoor acoustic scenes, 
such as bus, cafe, car, city center, forest path, library, train, etc. In total, 13 hours of stereo audio recordings are available, making this one of the largest datasets available. 

We perform experiments on six sets of features, including standard Mel-frequency cepstral coefficients (MFCC), Binaural MFCC, log Mel-spectrum and two different large-scale temporal pooling features extracted using OpenSMILE.
On these features, we apply five models: Gaussian Mixture Model (GMM), Deep Neural Network (DNN), Recurrent Neural Network (RNN), Convolutional Deep Neural Network (CNN) and i-vector. Using the late-fusion approach, we improve the performance of the baseline 72.5\% by 15.6\% in 4-fold Cross Validation (CV) avg. accuracy and 11\% in test accuracy, which matches the best result of the DCASE 2016 challenge. 

With large feature sets, deep neural network models outperform traditional methods and achieve the best performance among all the studied methods. Consistent with other work, the best performing single model is the non-temporal DNN model, which we take as evidence that sounds in the DCASE challenge do not exhibit strong temporal dynamics.
\end{abstract}
\vspace{-0.4cm}
\section{Introduction}
\label{sec:intro}
\vspace{-0.2cm}

Increasingly, machines in various environments can hear, such as smartphones, security systems, and autonomous robots. The prospect of human-like sound understanding could open up a wide range of applications, including intelligent machine state monitoring using acoustic information, acoustic surveillance, cataloging and information retrieval applications such as search in audio archives~\cite{r04} as well as audio-assisted multimedia content search. Compared with speech, environmental sounds are more diverse and span a wide range of frequencies. Moreover, they are often less well defined. Existing works for this task largely use conventional classifiers such as GMM and SVM, which do not have the feature abstraction capability found in deeper models. Furthermore, conventional models do not model temporal dynamics. For example, the winning solutions by~\cite{roma2013recurrence}\cite{Eghbal-Zadeh2016} for DCASE challenge 2013 and 2016, extracts MFCC and i-vectors, and they both used other deeper models for temporal relation analysis. In this work, we focus on the task of acoustic scene identification, which aims to characterize the acoustic environment of an audio stream by selecting a semantic label for it. 
We apply state-of-the-art deep learning (DL) architectures to various feature representations generated from signal processing methods. Specifically, we use the following architectures: (1) Deep Neural Network (DNN) (2) Recurrent Neural Network (RNN); (3) Convolutional Deep Neural Network (CNN). Additionally, we explore the combination of these models: (DNN, RNN and CNN). We also compare DL models with Gaussian mixture model (GMM), and i-vectors.\\
We also use several feature representations based on signal processing methods: Mel-frequency cepstral coefficients (MFCC), log Mel-Spectrum, spectrogram, other conventional features such as pitch, energy, zero-crossing rate, mean-crossing rate etc. There are several studies using DL in sound event detection~\cite{c15}\cite{m15}. However, to the best of our knowledge, this is the first \textit{comprehensive study} of a diverse set of deep architectures on the acoustic scene recognition task, borrowing ideas from signal processing as well as recent advancements in automatic speech recognition.
We use the dataset from the DCASE challenge. The dataset contains 15 diverse indoor and outdoor locations (classes), such as bus, cafe, car, city center, forest path, library, train, totaling 13 hours of audio recording (see Section 3.1 for detail). 
In this paper, we present a comparison of the most successful and complementary approaches to sound event detection on DCASE, which we implemented on top of our evaluation system~\cite{Dai2016} in a systematic and consistent way.
\vspace{-0.5cm}

\section{Experiments}
\label{sec:format}
\vspace{-0.3cm}

\subsection{Dataset}
\label{sec:pagestyle}
\vspace{-0.2cm}
We use the dataset from the IEEE challenge on Detection and Classification of Acoustic Scenes and Events~\cite{Heittola2016}, and we also use the evaluation setup from the contest. The training dataset contains 15 diverse indoor and outdoor locations (labels), totaling 9.75 hours of recording (1170 files) and 8.7GB in WAV format (Dual Channel, Sample Rate: 44100 Hz, Precision: 24 bit, Duration: 30 sec each). We do 4-fold CV for model selection and parameter tuning. The evaluation dataset (390 files) contains same classes of audio as training set, with totaling 3.25 hours of recording and 2.5GB in the same WAV format.

\vspace{-0.5cm}
\subsection{Features}
\vspace{-0.2cm}
\label{sec:typestyle}
We create six sets of features using audio signal processing methods:
\vspace{-0.2cm}
\begin{enumerate}[wide, labelwidth=!, labelindent=0pt]
    \item \emph{Monaural and Binaural MFCC}: Same as the winning solution in the DCASE challenge 2016~\cite{Eghbal-Zadeh2016}. We take 23 Mel-frequency (excluding the 0th) cepstral coefficients over window length 20 ms. We augment the feature with first and second order differences using 60 ms window, resulting in a 61-dimension vector. We also computed the MFCC on right, left and the channel difference (BiMFCC).
    \vspace{-0.2cm}
    \item \emph{Smile983 \& Smile6k}: We use OpenSmile~\cite{Eyben:2010:OMV:1873951.1874246} to generate MFCC, Fourier transforms, zero crossing rate, energy, and pitch, among others. We also compute first and second order features resulting in 6573 features. We select 983 features recommended by domain experts to create the 983-dim feature. Note that this is a much larger feature set than the MFCC features and each feature represents longer time window of 100 ms.
    \vspace{-0.2cm}
    \item \emph{LogMel}: We use LibROSA~\cite{brian_mcfee_2015_32193} to compute the log Mel-Spectrum, and we use the same parameters as the MFCC setup. This is the mel log powers before the discrete cosine transform step during the MFCC computation. We take 60 mel frequencies and 200 mel frequencies resulting in 60-dim and 200-dim LogMel features.
\end{enumerate}
\vspace{-0.2cm}
All features are standardized to have zero mean and unit variance on the training set. The same standardization is applied at the validation and test time. 

\vspace{-0.5cm}
\subsection{Models and Hyperparameter Tuning}
\label{sec:majhead}
\vspace{-0.2cm}
\subsubsection{Guassian Mixture Models (GMMs)}
\label{ssec:subhead}
\vspace{-0.2cm}
We use the GMMs provided by the DCASE challenge committee~\cite{Heittola2016} as the baseline system for acoustic scene recognition. Each audio clip is represented as a bag of acoustic features extracted from audio segments, and for each class label, a GMM is trained on this bag of acoustic features using only audio clips from that class. 

\vspace{-0.5cm}
\subsubsection{i-vector Pipeline}
\label{ssec:subhead}
\vspace{-0.2cm}
We replicate the i-vector~\cite{kenny2005eigenvoice} pipeline from~\cite{Eghbal-Zadeh2016}. The universal background model (UBM) is a GMM with 256 components trained on the development dataset using BiMFCC feature. The mean supervector $M$ of the GMM can be decomposed as: $M=m+T$$\cdot$$y$, where $m$ is an audio scene independent vector and $T$$\cdot$$y$ is an offset. The low-dimensional (400-dim) subspace vector $y$ is an audio scene dependent vector, and it is a latent variable with the normal prior. The i-vector $w$ is a maximum a posteriori probability (MAP) estimate of $y$. We use the Kaldi Toolkit~\cite{Povey11thekaldi} to compute $T$ matrix and perform Linear Discrimant Analysis (LDA).

\vspace{-0.5cm}
\subsubsection{Deep Neural Networks (DNNs)}
\label{ssec:subhead}
\vspace{-0.2cm}
Multi-layer perception has recently been successfully applied to speech recognition and audio analysis and shows superior performance compared to GMMs~\cite{DBLP:journals/corr/abs-1303-5778}. Here we tried various sets of hyperparameters including depth (2-10 layers), number of hidden units (256-1024), dropout rate (0-0.4), regularizer (L1, L2), and various optimization algorithms(stochastic gradient descent, Adam~\cite{DBLP:journals/corr/KingmaB14}, RMSProp~\cite{dauphin2015rmsprop}, Adagrad~\cite{dhs11}), batch normalization~\cite{is15}, etc. All the deep models we tried in the next two sections are tuned via cross validation (CV) to achieve their best performance.

\vspace{-0.5cm}
\subsubsection{Recurrent Neural Networks (RNNs)}
\label{sssec:subsubhead}
\vspace{-0.2cm}

Bidirectional architectures generally perform better than uni-directional counterparts. We tried both LSTM~\cite{Hochreiter:1997:LSM:1246443.1246450} and GRU~\cite{DBLP:journals/corr/ChungGCB14} bidirectional layers. Our network only has 2 layers (one direction a layer) due to convergence time and limited improvement from deeper RNN models~\cite{irsoyopinion}.

\vspace{-0.5cm}
\subsubsection{Convolutional Neural Networks (CNNs)}
\label{sssec:subsubhead}
\vspace{-0.2cm}
 Lately, CNNs has been applied to speech recognition using spectrogram features~\cite{DBLP:journals/corr/HannunCCCDEPSSCN14} and achieve state-of-the-art speech recognition performance. We employ architectures similar to the VGG net~\cite{DBLP:journals/corr/SimonyanZ14a} to keep the number of model parameters small. The input  we use is the popular rectified linear units (relu) to model non-linearity. We also found that dense layers in the bottom do not help but only slow down computation, so we do not include them in most experiments. Dropout layers significantly improve performance, which is consistent with the CNN behaviors on natural images. Overall CNNs take significantly longer to train than RNNs, and DNNs due to the convolutional layers. \\
 Table 1 shows an example of the architectures of all the DL models described as above.\\
 
\hspace*{-0.6cm} 
\begin{minipage}[l]{0.5\textwidth}
  \begin{tabular}{ |c|c|c|}
    \hline
    \bf{DNN} & \bf{RNN} & \bf{CNN} \\ \hline
    \multicolumn{3}{|c|}{Input depending on feature} \\ \hline
    Dense 256 & GRU 256& 32$\times$3$\times$3-BN-ReLu \\ \hline
    BN + Dropout0.2 & GRU 256& 32$\times$3$\times$3-BN-ReLu \\ \hline
    Dense 256 & Dropout0.4 & MaxPool2$\times$2+Dropout0.3 \\ \hline
    BN + Dropout0.2 & BN & 64$\times$3$\times$3-BN-ReLu \\ \hline
    Dense 256 &  & 64$\times$3$\times$3-BN-ReLu\\ \hline
    BN + Dropout0.2 &  & MaxPool2$\times$2+Dropout0.3\\ \hline
    Dense 256 & & 128$\times$3$\times$3-BN-ReLu \\ \hline
    BN + Dropout0.2&  & 128$\times$3$\times$3-BN-ReLu\\ \hline
     &  & MaxPool2$\times$2+Dropout0.3\\ \hline
    \multicolumn{3}{|c|}{15-way Softmax} \\ \hline
  \end{tabular}
  
  \centerline{\textit{\small Table 1: Model Specifications. BN:Batch Normalization}}
  \centerline{\textit{\small ReLu: Rectified Linear Activation Function}} 
\end{minipage}

\vspace{-0.3cm}
\subsection{Pipeline \& System Configuration}
\label{sssec:subsubhead}
\vspace{-0.2cm}
For each audio clip (train and test), our processing pipeline consists of the following: 1) Apply the various transforms (Section 2.2) to each audio clip to extract the feature representations; 2) For non-temporal models such as GMMs, we treat each feature as a training example. For temporal models such as RNNs, we consider a sequence of features as one training example; 3) At test time, we apply the same pipeline as training and break the audio clip as multiple instances, and the likelihood of a class label for a test audio clip is the sum of predicted class likelihood for each segment. The class with the highest predicted likelihoods is the predicted label for the test audio clip.\\
We train our deep learning models with the Keras library~\cite{c15} built on Theano~\cite{theano} and TensorFlow, using 4 Titan X GPUs on a 128GB memory, Intel Core i7 node.

\vspace{-0.5cm}
\subsection{Late Fusion}
\label{sssec:subsubhead}
\vspace{-0.2cm}
In the end, we ensemble all the models mentioned above. In total, we have thirty models for the problem and five different architectures. We rank the models by performance, only best performing models which pass a predefined accuracy threshold are included in fusion. To further stabilize the model, we construct ensembles of the ensembles. For example, the baseline GMM is excluded due to its poor performance. We test with random forest, extremely randomized trees, Ada-boost, gradient tree boosting, weighted average probabilities and other model selection methods in the late fusion~\cite{c04}.

\vspace{-0.6cm}
\section{Results}
\label{sec:print}
\vspace{-0.2cm}
Figure 1 shows the cross validation (CV) accuracy for 5 classifiers over 6 features. 60-dim and 200-dim LogMel are listed in a single column. GMM with MFCC feature is the official baseline provided in the DCASE challenge, which achieves a mean CV accuracy of 72.5\%, while our best performing model (DNN with the Smile6k features) achieves a mean CV accuracy of 84.2\% and test accuracy of 84.1\%.
The best late fusion model has an 88.1\% mean CV accuracy and 88.2\% test accuracy, which is competitve with the winning solution in the DCASE challenge~\cite{Eghbal-Zadeh2016}.

\hspace*{-1.4cm}
\begin{tikzpicture}
\begin{axis}[
    width=10.2cm, height= 7.0cm,
    xtick={0.8,...,4,4.5},
    xticklabels={%
        MFCC,
        BiMFCC,
        Smile983,
        Smile6k,
        LogMel},
    grid=major,
    ybar=1*\pgflinewidth,
    bar width=7pt,
    enlarge x limits=0.15,
    legend style={at={(0.5,0.95)}, anchor=south,legend columns=-1}
    ]

\addplot[
    fill=bblue,
    draw=black,
    point meta=y,
    every node near coord/.style={inner ysep=5pt},
    error bars/.cd,
        y dir=both,
        y explicit
]coordinates{
        (1,0.7252) += (0,-0.095) -=(0,-0.0535)
        (2,0.7697) +=(0,-0.080) -=(0,-0.0392)
        (3,0.6893) +=(0,-0.0488) -= (0,-0.0376)
        (4,0.6027) +=(0,-0.0587) -= (0,-0.0320)
        };

\addplot[
    fill=ggreen,
    draw=black,
    point meta=y,
    every node near coord/.style={inner ysep=5pt},
    error bars/.cd,
        y dir=both,
        y explicit
] coordinates{
        (1,0.7522) += (0,-0.090) -=(0,-0.0535)
        (2,0.8171) +=(0,-0.0320) -=(0,-0.0392)
        (3,0.7801) +=(0,-0.0788) -= (0,-0.0476)
        (4,0.7724) +=(0,-0.0887) -= (0,-0.0520)
        };

\addplot[
    fill=rred,
    draw=black,
    point meta=y,
    every node near coord/.style={inner ysep=5pt},
    error bars/.cd,
        y dir=both,
        y explicit
] coordinates{
        (1,0.7022) += (0,-0.030) -=(0,-0.0535)
        (2,0.7770) +=(0,-0.0620) -=(0,-0.0292)
        (3,0.7891) +=(0,-0.0388) -= (0,-0.0576)
        (4,0.8424) +=(0,-0.0687) -= (0,-0.0520)
        };
        
\addplot[
    fill=ppurple,
    draw=black,
    point meta=y,
    every node near coord/.style={inner ysep=5pt},
    error bars/.cd,
        y dir=both,
        y explicit
] coordinates{
        (1,0.7322) += (0,-0.030) -=(0,-0.0535)
        (2,0.7870) +=(0,-0.0620) -=(0,-0.0292)
        (3,0.7991) +=(0,-0.0488) -= (0,-0.0476)
        (4,0.8024) +=(0,-0.0687) -= (0,-0.0520)
        };
        
\addplot[
    fill=red!,
    draw=black,
    point meta=y,
    every node near coord/.style={inner ysep=5pt},
    error bars/.cd,
        y dir=both,
        y explicit
] coordinates{
        (1,0.7522) += (0,-0.030) -=(0,-0.0535)
        (2,0.7870) +=(0,-0.0620) -=(0,-0.0292)
        (4.3,0.822) +=(0,-0.0620) -=(0,-0.0292)
        (3,0.7791) +=(0,-0.0888) -= (0,-0.0576)
        (4,0.7924) +=(0,-0.0687) -= (0,-0.0320)
        };
          
\addplot[
    fill=yellow!,
    draw=black,
    point meta=y,
    every node near coord/.style={inner ysep=5pt},
    error bars/.cd,
        y dir=both,
        y explicit
] coordinates{
        (4.7,0.881) +=(0,-0.0320) -=(0,-0.0422)
        };      
\legend{GMM,IVector,DNN,RNN,CNN,LateFusion}
\end{axis}
\end{tikzpicture}
\centerline {\textit{\small Figure 1: 4-fold CV avg. accuracy}}

\hspace*{-0.85cm}
\begin{tikzpicture}
\matrix (first) [table,text width=6ex]
{
& GMM   & I-Vector & DNN & RNN & CNN & Fusion\\
Beach   & 69.3 & 80.7 & 89.8 & 80.3 & 78.7 & 92.3 \\
Bus     & 79.6 & 82.4 & 95.3 & 88.6 & 72.1 & 95.3\\
Cafe/Rest. & 83.2 & 70.0 & 69.9 & 64.7 & 66.4 & 79.9\\
Car     & 87.2 & 96.1 & 87.2 & 88.8 & 99.1 & 97.2\\
City    & 85.5 & 90.0 & 97.3 & 96.2 & 93.5 & 89.2\\
Forest  & 81.0 & 92.0 & 96.4 & 95.0 & 99.8 & 99.8\\
Grocery & 65.0 & 93.8 & 79.3 & 75.5 & 85.3 & 96.2\\
Home    & 82.1 & 65.2 & 84.8 & 75.7 & 82.9 & 88.2\\
Library & 50.4 & 76.1 & 81.2 & 81.6 & 72.7 & 86.2\\
Metro   & 94.7 & 83.5 & 97.3 & 93.7 & 98.7 & 92.3\\
Office  & 98.6 & 93.1 & 99.7 & 79.6 & 97.6 & 99.7\\
Park    & 13.9 & 78.6 & 49.4 & 45.8 & 45.7 & 71.2\\
Resident& 77.7 & 66.5 & 76.9 & 68.7 & 81.6 & 77.0\\
Train   & 33.6 & 72.4 & 51.1 & 61.2 & 59.2 & 65.2\\
Tram    & 85.4 & 84.6 & 97.0 & 90.7 & 91.7 & 92.2\\
Average & 72.5 & 81.7 & 84.2 & 80.2 & 82.2 & 88.1\\
};
\end{tikzpicture}
\textit{\small Table 2: Class-wise accuracy (\%) of the best CV average models. Colored rows correspond to the most challenging classes in the confusion matrix from~\cite{Dai2016}}

\vspace{-0.6cm}
\section{Discussion}
\label{sec:page}
\vspace{-0.2cm}

Figure 1 shows that feature representation is critical for classifier performance. For neural network models (RNNs, DNNs), a larger set of features extracted from signal processing pipeline improves performance. Among the neural network models, it is interesting to note that RNNs and CNNs outperform DNNs using MFCC, BiMFCC and Smile983 features, but DNNs outperform RNNs and CNNs on Smile6k feature.
It is possible that with limited feature representation (e.g., MFCC and BiMFCC), modeling temporally adjacent pieces enhances the local feature representation and thus improves performance of temporal models like RNNs. However, with a sufficiently expressive feature (e.g., Smile6k), the temporal modeling becomes less important, and it becomes more effective to model local dynamics rather than long-range dependencies.
Unlike speech, which has long range dependency (a sentence utterance could span 6-20 seconds), environment sounds generally lack a coherent context, as events in the environment occur more or less randomly from the listener's perspective. A human listener of environmental noise is unlikely able to predict what sound will occur next in an environment, in contrast to speech. \\ 
Table 2 shows that most locations are relatively easy to identify except for a few difficult classes to distinguish, such as park and residential area, or train and tram. We can also see that various models have varying performance in different classes, and thus performing late fusion tends to compensate individual model's error, leading to improved overall performance. 

\vspace{-0.5cm}
\subsection{GMM \& I-Vectors}
\vspace{-0.2cm}
The performance of non-neural network models, particularly, the GMMs suffer from the \emph{curse of dimensionality}. That is, in the high dimensional space, the volume grows exponentially while the number of available data stays constant, leading to highly sparse sample. In spite of being GMM-based, this issue is less prominent for the i-vector approach since its Factorial Analysis procedure always keep the dimension low. The performance of i-vector pipeline is the best among all the models using BiMFCC feature, we observe i-vector pipeline outperform DL models with low dimension features. We also observe i-vector pipeline tends to do better in more noisy classes such as train and tram, while suffer in relatively quiet classes such as home and residential area.

\vspace{-0.5cm}
\subsection{DNN}
\vspace{-0.2cm}  
Deep classifiers are able to learn more abstract representation of the feature. Figure 2 shows the 1st layer of a fully connected DNN model. Here, our feature is BiMFCC (61-dim). The DNN model has 5 dense layers, each with 256 hidden units. Figure2(a) shows the FFT of the weight of the first layer, and indicates the responsiveness of the 256 corresponding hidden units. We note that DNN's neurons are more active in the MFCC range (0-23) and are less active in the delta of MFCC (24-41) and double delta dimension (42-61). If we apply a Savitzky-Golay smoothing function~\cite{citeulike:4226570} which acts like a low-pass filter on each neuron's vector (61-dim). We get Figure2(b) which is the de-noised weight of layer (each colored line corresponds with one neuron vector), which looks like a filter bank. The chaotic responses of DNN neurons also demonstrate that DNN is not capable of capturing temporal information in the feature. 

\begin{figure}
\begin{minipage}[b]{.48\linewidth}
  \centering
  \centerline{\includegraphics[width=4.8cm, height=3.0cm]{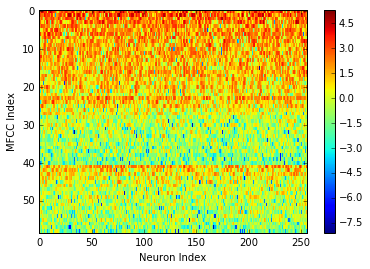}}
  \centerline{\textit{\small (a) Weight after FFT}}\medskip
\end{minipage}
\hfill
\begin{minipage}[b]{0.48\linewidth}
  \centering
  \centerline{\includegraphics[width=4.0cm, height=3.0cm]{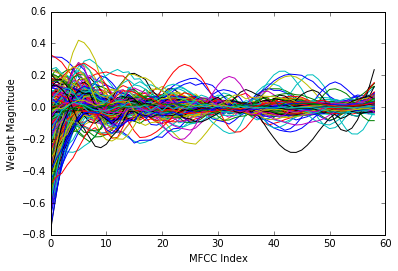}}
  \centerline{\textit{\small (b) Weight after Smoothing}}\medskip
\end{minipage}
  \centerline{\textit{\small Figure 2: DNN's 1st layer after input}}
  \vspace{-0.9cm}

\end{figure}

\begin{figure}[htb]
  \vspace{-0.2cm}
  \centering
  \centerline{\includegraphics[width=10cm,height=2.6cm]{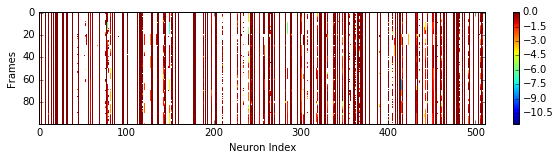}}
  \centerline{\textit{\small Figure 3: RNN Neuron (512-dim) Activation} }
  \centering
  \centerline{\includegraphics[width=10cm,height=2.6cm]{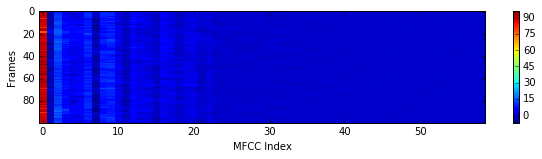}}
  \centerline{\textit{\small Figure 4: BiMFCC 61 over 100 frames}}
\end{figure}
\vspace{-1.0cm}

\subsection{RNN}
\vspace{-0.2cm}
Our RNN model consists of 2 bidirectional GRU layers, and they both have 512 hidden units. Figure 3 shows the neuron activation of the forward layer of the bidirectional GRU network over 100 frames. Figure 4 shows the corresponding input feature (MFCC). With a train audio, it shows that RNN neurons are stable across the time domain as long as there is no variation of feature over time. This shed light on why our RNN performs better on relatively more monotonous audio scenes such as train and tram rather than event-rich audio scenes like park and residential areas. Meanwhile, there could be a potential gain from incorporating attention-based RNN~\cite{DBLP:journals/corr/ChorowskiBCB14} here to tackle those event-rich audio scenes based on audio events. 


\vspace{-0.5cm}
\subsection{CNN}
\begin{figure}
\begin{minipage}[b]{.48\linewidth}
  \centering
  \centerline{\includegraphics[width=4.2cm, height=3cm]{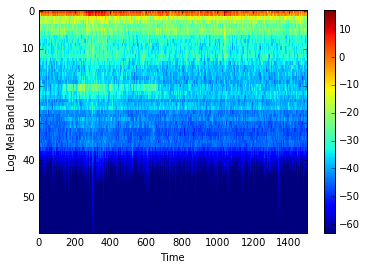}}
  \centerline{\textit{\small (a)Log Mel-spectrum}}\medskip
\end{minipage}
\hfill
\begin{minipage}[b]{0.48\linewidth}
  \centering
  \centerline{\includegraphics[width=4.9cm, height=3cm]{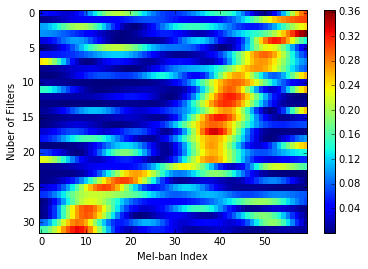}}
  \centerline{\textit{\small (b)Weight after FFT}}\medskip
\end{minipage}
\vspace{-0.6cm}
\centerline{\textit{\small Figure 5: CNN 1st Convolutional2D layer}}
\end{figure}
\vspace{-0.2cm}
Figure 5(a) shows the input to the CNN which is a log Mel energy spectrum (60-dim). Figure 5(b) is the weight of the 1st convolutional layer (32 convolutional filters) after FFT. This highly resembles a filter bank of bandpass filters. We notice there is a sharp transition in filters at around the 40th Mel band. This is due to the weak energy beyond the 40th Mel band shown in Figure 5(a). Our finding is consistent with prior work on speech data~\cite{golik2015convolutional}. The filter bank we learned are relatively wider compared with that is learned in speech. 


\vspace{-0.3cm}
\section{Conclusion}
\label{sec:foot}
\vspace{-0.2cm}
We find that deep learning models compare favorably with traditional pipelines (GMM and i-vector). Specifically, GMMs with the MFCC feature, the baseline model provided by DCASE contest, achieves 77.2\% test accuracy, while the best performing model (hierarchical DNN with Smile6k feature) reaches 88.2\% test accuracy. RNN and CNN generally have performance in the range of 73-82\%. \\
Fusing the temporal specialized models (e.g. CNNs, RNNs) with resolution specialized models (DNNs, i-vector) improve the overall performance significantly. We train the classifiers independently first to maximize model diversity, and fuse these models for the best performance.
We find that no single model outperforms all other models across all feature sets, showing that model performance can vary significantly with feature representation. The fact that the best performing model is the non-temporal DNN model is evidence that environmental (or ``scene'') sounds do not necessarily exhibit strong temporal dynamics. This is consistent with our day-to-day experience that environmental sounds tend to be random and unpredictable.


%
%


\vfill\pagebreak

\bibliographystyle{IEEEbib}

\bibliography{Master}

\end{document}